\documentclass{article}%
\usepackage{amsfonts}
\usepackage{amsmath}
\usepackage{amssymb}
\usepackage{graphicx}%
\providecommand{\U}[1]{\protect\rule{.1in}{.1in}}

\begin{document}

\title{Generalized supersymmetric cosmological term in N=1 Supergravity}
\author{P. K. Concha$^{1,2,3}$, E. K. Rodr\'{\i}guez$^{1,2,3}$, P. Salgado$^{1}$\\$^{1}${\small {\textit{Departamento de F\'{\i}sica, Universidad de
Concepci\'{o}n,}} }\\{\small {\textit{Casilla 160-C, Concepci\'{o}n, Chile.}} }\\\ $^{2}${\small {\textit{Dipartimento di Scienza Applicata e Tecnologia
(DISAT),}} }\\{\small {\textit{Politecnico di Torino, Corso Duca degli Abruzzi 24,}}}\\{\small {\textit{I-10129 Torino, Italia.}} }\\\ $^{3}${\small {\textit{Istituto Nazionale di Fisica Nucleare (INFN) Sezione
di Torino,}} }\\{\small {\textit{Via Pietro Giuria 1, 10125 Torino, Italia.}}}\\
\\{\small {\textit{E-mail:} {\texttt{patrickconcha@udec.cl},
\texttt{everodriguez@udec.cl}, \texttt{pasalgad@udec.cl}}}}}
\maketitle

\begin{abstract}
An alternative way of introducing the supersymmetric cosmological term in a
supergravity theory is presented. \ We show that the $AdS$-Lorentz
superalgebra allows to construct a geometrical formulation of supergravity
containing a generalized supersymmetric cosmological constant. The $N=1$,
$D=4$ supergravity action is built only from the curvatures of the
$AdS$-Lorentz superalgebra and corresponds to a MacDowell-Mansouri like
action. The extension to a generalized $AdS$-Lorentz superalgebra is also analyzed.

\end{abstract}

\section{Introduction}

\qquad A good candidate to describe the dark energy corresponds to the
cosmological constant \cite{FTH, Padmanabhan}. \ It is well known that a
cosmological term can be introduced in a $D=4$ gravity theory using the Anti
de Sitter ($AdS$) algebra. \ In particular the supersymmetric extension of
gravity including a cosmological term can be obtained in a geometric
formulation. \ In this framework, supergravity is built from the curvatures of
the $\mathfrak{osp}\left(  4|1\right)  $ superalgebra and the action is known
as the MacDowell-Mansouri action \cite{MM}.

Recently it was presented in ref. \cite{AKL} an alternative way of introducing
the generalized cosmological constant term using the Maxwell algebra. \ It is
usually accepted that the symmetries of Minkowski spacetime are described by
the Poincar\'{e} algebra. In refs. \cite{BCR, Schrader} this spacetime was
generalized extending its symmetries from the Poincar\'{e} to the Maxwell
symmetries whose generators satisfy the following commutation relations%
\begin{align}
\left[  P_{a},P_{b}\right]   &  =\Lambda Z_{ab},\text{ \ \ }Z_{ab}%
=-Z_{ba}\text{ ,}\\
\left[  J_{ab},Z_{cd}\right]   &  =\eta_{bc}Z_{ad}-\eta_{ac}Z_{bd}-\eta
_{bd}Z_{ac}+\eta_{ad}Z_{bc},\\
\left[  Z_{ab},Z_{cd}\right]   &  =0,\text{ \ \ \ \ }\left[  Z_{ab}%
,P_{c}\right]  =0.
\end{align}
Here $Z_{ab}$ correspond to tensorial Abelian charges and the constant
$\Lambda$ can be related to the cosmological constant when $\left[
{\Large \Lambda}\right]  =M^{2}$. \ If we put $\Lambda=e$, where $e$ is the
electromagnetic coupling constant, we have the possible description of
spacetime in presence of a constant electromagnetic background field.

The deformations of the Maxwell symmetries lead to the $\mathfrak{so}%
(D-1,2)\oplus\mathfrak{so(}D-1,1)$ or $\mathfrak{so}(D,1)\oplus\mathfrak{so}%
(D-1,1)$ algebra \cite{DKGS, Sorokas}. In this case the $Z_{ab}$ generators
are non-abelian. If spacetime symmetries are considered local symmetries then
it is possible to construct Chern-Simons gravity actions where dark energy
could be interpreted as part of the metric of spacetime.

Subsequently it was shown in ref. \cite{SS} that the generalized cosmological
constant term can also be included in a Born-Infeld like action built from the
curvatures of the $AdS$-Lorentz%
\footnote{Also known as $\mathfrak{so}(D-1,1)\oplus\mathfrak{so}%
(D-1,2)$ algebra.}
($AdS-\mathcal{L}_{4}$) algebra. \ Alternatively the $AdS$-Lorentz algebra can
be obtained as an abelian semigroup expansion ($S$-expansion) of the $AdS$
algebra using $S_{\mathcal{M}}^{\left(  2\right)  }$ as the relevant semigroup
\cite{DFIMRSV}.

The $S$-expansion procedure is based on combining the multiplication law of a
semigroup $S$ with the structure constants of a Lie algebra $\mathfrak{g}$
\cite{Sexp}. \ The new Lie algebra obtained using this method is called the
$S$-expanded algebra $\mathfrak{G}=S\times\mathfrak{g}$.\ \ Diverse
(super)gravity theories have been extensively studied using the $S$-expansion
approach. \ In particular, interesting results have been obtained in refs.
\cite{IRS1, GRCS, CPRS1, GSRS, Topgrav, BDgrav, FISV, CPRS2, CR2, CPRS3}. \ An
alternative expansion method can be found in ref. \cite{AIPV}.

In this paper we analyze the consequence of considering the supersymmetric
extension of the $AdS$-Lorentz algebra in the construction of a\ supergravity
theory.\ This superalgebra has the following anticommutation relation,%
\begin{equation}
\left\{  Q_{\alpha},Q_{\beta}\right\}  =-\frac{1}{2}\left[  \left(
\gamma^{ab}C\right)  _{\alpha\beta}Z_{ab}-2\left(  \gamma^{a}C\right)
_{\alpha\beta}P_{a}\right]  ,
\end{equation}
where $Q_{\alpha}$ represents a $4$-component Majorana spinor charge. \ Unlike
the Maxwell superalgebra the new generators $Z_{ab}$ are not abelian and
behave as a Lorentz generator,%
\begin{equation}
\left[  Z_{ab},Z_{cd}\right]  =\eta_{bc}Z_{ad}-\eta_{ac}Z_{bd}-\eta_{bd}%
Z_{ac}+\eta_{ad}Z_{bc}.
\end{equation}
The presence of the $Z_{ab}$ generators implies the introduction of a new
bosonic "matter" field $k^{ab}$ which modifies the definition of the different
curvatures. \ In particular, we are interested in studying the geometrical
consequences of including the generators $Z_{ab}=\left[  P_{a},P_{b}\right]  $
in supergravity. \ Although the same non-commutativity is present in the
Maxwell symmetries, it was shown in ref. \cite{CR2} that the supergravity
action \`{a} la MacDowell-Mansouri based on the Maxwell superalgebra does not
reproduce the cosmological constant term in the action. \ Then, the
$AdS$-Lorentz superalgebra seems to be a better candidate in order to
introduce the cosmological term in supergravity, in presence of the bosonic
generators $Z_{ab}$.

On the other hand, as shown in ref. \cite{MO, MTO} the four-dimensional
renormalized action for $AdS$ gravity, which corresponds to the bosonic
MacDowell-Mansouri action, is equivalent on-shell to the square of the Weyl
tensor describing conformal gravity. \ Then, the supergravity action \`{a} la
MacDowell-Mansouri suggests a superconformal structure which represents an
additional motivation in our construction.

It is the purpose of this work to construct a supergravity action which
contains a generalized supersymmetric cosmological constant from the
$AdS$-Lorentz superalgebra. \ To this aim, we apply the $S$-expansion method
to the $\mathfrak{osp}\left(  4|1\right)  $ superalgebra and we build a
MacDowell-Mansouri like action with the expanded $2$-form curvatures. \ \ The
result presented here corresponds to an alternative way of introducing the
supersymmetric cosmological term and can be seen as the supersymmetric
extension of refs. \cite{AKL, SS}. \ We extend our result introducing the
generalized minimal $AdS$-Lorentz superalgebra and we build a more general
$D=4$, $N=1$ supergravity action involving a supersymmetric cosmological term.

This work is organized as follows: in section 2 we review the construction of
the $AdS$-Lorentz superalgebra using the $S$-expansion procedure. \ Sections 3
and 4 contain our main results. \ In section 3, we present the $D=4$, $N=1$
supergravity action including a generalized supersymmetric cosmological
constant. \ We show that this action corresponds to a MacDowell-Mansouri like
action built from the curvatures of the $AdS$-Lorentz superalgebra. \ In
section 4 we extend our results to the generalized minimal $AdS$-Lorentz
superalgebra. \ Section 5 concludes the work with some comments about possible
development and usefulness of our results.

\section{$AdS-$Lorentz superalgebra and the abelian semigroup expansion
procedure}

\qquad The abelian semigroup expansion procedure ($S$-expansion) is a powerful
tool in order to derive new Lie (super)algebras \cite{Sexp}. \ Furthermore,
the $S$-expansion method has the advantage to provide with an invariant tensor
for the $S$-expanded algebra $\mathfrak{G}=S\times\mathfrak{g}$ in terms of an
invariant tensor for the original algebra $\mathfrak{g}$.

Following refs. \cite{Sexp, FISV}, it is possible to obtain the $AdS-$Lorentz
superalgebra as an $S$-expansion of the $\mathfrak{osp}\left(  4|1\right)  $
superalgebra using $S_{\mathcal{M}}^{\left(  2\right)  }$ as the abelian semigroup.

Before applying the $S$-expansion method it is necessary to consider a
decomposition of the original algebra $\mathfrak{g}=\mathfrak{osp}\left(
4|1\right)  $ in subspaces $V_{p}$,%
\begin{align}
\mathfrak{g}=\mathfrak{osp}\left(  4|1\right)   &  =\mathfrak{so}\left(
3,1\right)  \oplus\frac{\mathfrak{osp}\left(  4|1\right)  }{\mathfrak{sp}%
\left(  4\right)  }\oplus\frac{\mathfrak{sp}\left(  4\right)  }{\mathfrak{so}%
\left(  3,1\right)  }\nonumber\\
&  =V_{0}\oplus V_{1}\oplus V_{2},
\end{align}
where $V_{0}$ is generated by the Lorentz generator $\tilde{J}_{ab}$, $V_{1}$
corresponds to the fermionic subspace generated by$\ $a $4$-component Majorana
spinor charge $\tilde{Q}_{\alpha}$ and $V_{2}$ corresponds to the $AdS$ boost
generated by $\tilde{P}_{a}$. The $\mathfrak{osp}\left(  4|1\right)  $
generators satisfy the following (anti)commutation relations%
\begin{align}
\left[  \tilde{J}_{ab},\tilde{J}_{cd}\right]   &  =\eta_{bc}\tilde{J}%
_{ad}-\eta_{ac}\tilde{J}_{bd}-\eta_{bd}\tilde{J}_{ac}+\eta_{ad}\tilde{J}%
_{bc},\label{ads1}\\
\left[  \tilde{J}_{ab},\tilde{P}_{c}\right]   &  =\eta_{bc}\tilde{P}_{a}%
-\eta_{ac}\tilde{P}_{b},\\
\left[  \tilde{P}_{a},\tilde{P}_{b}\right]   &  =\tilde{J}_{ab},\\
\left[  \tilde{J}_{ab},\tilde{Q}_{\alpha}\right]   &  =-\frac{1}{2}\left(
\gamma_{ab}\tilde{Q}\right)  _{\alpha},\text{ \ \ \ \ }\left[  \tilde{P}%
_{a},\tilde{Q}_{\alpha}\right]  =-\frac{1}{2}\left(  \gamma_{a}\tilde
{Q}\right)  _{\alpha},\\
\left\{  \tilde{Q}_{\alpha},\tilde{Q}_{\beta}\right\}   &  =-\frac{1}%
{2}\left[  \left(  \gamma^{ab}C\right)  _{\alpha\beta}\tilde{J}_{ab}-2\left(
\gamma^{a}C\right)  _{\alpha\beta}\tilde{P}_{a}\right]  . \label{ads5}%
\end{align}
Here, $\gamma_{a}$ are Dirac matrices and $C$ stands for the charge
conjugation matrix.

The subspace structure may be written as%
\begin{equation}%
\begin{tabular}
[c]{ll}%
$\left[  V_{0},V_{0}\right]  \subset V_{0},$ & $\left[  V_{1},V_{1}\right]
\subset V_{0}\oplus V_{2},$\\
$\left[  V_{0},V_{1}\right]  \subset V_{1},$ & $\left[  V_{1},V_{2}\right]
\subset V_{1,}$\\
$\left[  V_{0},V_{2}\right]  \subset V_{2},$ & $\left[  V_{2},V_{2}\right]
\subset V_{0}.$%
\end{tabular}
\label{subdes1}%
\end{equation}

Following the definitions of ref. \cite{Sexp}, let $S_{\mathcal{M}}^{\left(
2\right)  }=\left\{  \lambda_{0},\lambda_{1},\lambda_{2}\right\}  $ be an
abelian semigroup whose elements satisfy the multiplication law,%
\begin{equation}
\lambda_{\alpha}\lambda_{\beta}=\left\{
\begin{array}
[c]{c}%
\lambda_{\alpha+\beta},\text{ \ \ if }\alpha+\beta\leq2\\
\lambda_{\alpha+\beta-2},\text{ \ if }\alpha+\beta>2\text{\ }%
\end{array}
\right.  \label{lm1}%
\end{equation}
Let us consider the subset decomposition $S_{\mathcal{M}}^{\left(  2\right)
}=S_{0}\cup S_{1}\cup S_{2}$, with%
\begin{align}
S_{0}  &  =\left\{  \lambda_{0},\lambda_{2}\right\}  ,\\
S_{1}  &  =\left\{  \lambda_{1}\right\}  ,\\
S_{2}  &  =\left\{  \lambda_{2}\right\}  .
\end{align}
One sees that this decomposition is said to be resonant since it satisfies the
same structure as the subspaces $V_{p}$ [compare with eqs $\left(
\ref{subdes1}\right)  $]%
\begin{equation}%
\begin{tabular}
[c]{ll}%
$S_{0}\cdot S_{0}\subset S_{0},$ & $S_{1}\cdot S_{1}\subset S_{0}\cap S_{2}%
,$\\
$S_{0}\cdot S_{1}\subset S_{1},$ & $S_{1}\cdot S_{2}\subset S_{1},$\\
$S_{0}\cdot S_{2}\subset S_{2},$ & $S_{2}\cdot S_{2}\subset S_{0}.$%
\end{tabular}
\end{equation}
Following theorem IV.2 of ref. \cite{Sexp}, we can say that the superalgebra
\begin{equation}
\mathfrak{G}_{R}=W_{0}\oplus W_{1}\oplus W_{2}\text{,}%
\end{equation}
is a resonant subalgebra of $S_{\mathcal{M}}^{\left(  2\right)  }%
\times\mathfrak{g}$, where%
\begin{align}
W_{0}  &  =\left(  S_{0}\times V_{0}\right)  =\left\{  \lambda_{0},\lambda
_{2}\right\}  \times\left\{  \tilde{J}_{ab}\right\}  =\left\{  \lambda
_{0}\tilde{J}_{ab},\lambda_{2}\tilde{J}_{ab}\right\}  ,\\
W_{1}  &  =\left(  S_{1}\times V_{1}\right)  =\left\{  \lambda_{1}\right\}
\times\left\{  \tilde{Q}_{\alpha}\right\}  =\left\{  \lambda_{1}\tilde
{Q}_{\alpha}\right\}  ,\\
W_{2}  &  =\left(  S_{2}\times V_{2}\right)  =\left\{  \lambda_{2}\right\}
\times\left\{  \tilde{P}_{a}\right\}  =\left\{  \lambda_{2}\tilde{P}%
_{a}\right\}  .
\end{align}

Thus the new superalgebra is generated by $\left\{  J_{ab},P_{a}%
,Z_{ab},Q_{\alpha}\right\}  $, where these new generators can be written as%
\begin{align*}
J_{ab}  &  =\lambda_{0}\tilde{J}_{ab},\\
Z_{ab}  &  =\lambda_{2}\tilde{J}_{ab},\\
P_{a}  &  =\lambda_{2}\tilde{P}_{a},\\
Q_{\alpha}  &  =\lambda_{1}\tilde{Q}_{\alpha}.
\end{align*}
The expanded generators satisfy the (anti)commutation relations%
\begin{align}
\left[  J_{ab},J_{cd}\right]   &  =\eta_{bc}J_{ad}-\eta_{ac}J_{bd}-\eta
_{bd}J_{ac}+\eta_{ad}J_{bc},\label{ADSL401}\\
\left[  J_{ab},Z_{cd}\right]   &  =\eta_{bc}Z_{ad}-\eta_{ac}Z_{bd}-\eta
_{bd}Z_{ac}+\eta_{ad}Z_{bc},\\
\left[  Z_{ab},Z_{cd}\right]   &  =\eta_{bc}Z_{ad}-\eta_{ac}Z_{bd}-\eta
_{bd}Z_{ac}+\eta_{ad}Z_{bc},\\
\left[  J_{ab},P_{c}\right]   &  =\eta_{bc}P_{a}-\eta_{ac}P_{b},\text{
\ \ \ \ }\left[  P_{a},P_{b}\right]  =Z_{ab},\\
\left[  Z_{ab},P_{c}\right]   &  =\eta_{bc}P_{a}-\eta_{ac}P_{b},\\
\left[  J_{ab},Q_{\alpha}\right]   &  =-\frac{1}{2}\left(  \gamma
_{ab}Q\right)  _{\alpha},\text{ \ \ \ \ }\left[  P_{a},Q_{\alpha}\right]
=-\frac{1}{2}\left(  \gamma_{a}Q\right)  _{\alpha},\\
\left[  Z_{ab},Q_{\alpha}\right]   &  =-\frac{1}{2}\left(  \gamma
_{ab}Q\right)  _{\alpha},\\
\left\{  Q_{\alpha},Q_{\beta}\right\}   &  =-\frac{1}{2}\left[  \left(
\gamma^{ab}C\right)  _{\alpha\beta}Z_{ab}-2\left(  \gamma^{a}C\right)
_{\alpha\beta}P_{a}\right]  , \label{ADSL408}%
\end{align}
where we have used the multiplication law of the semigroup $\left(
\ref{lm1}\right)  $ and the commutation relations of the original
superalgebra. \ The new superalgebra obtained after a resonant $S_{\mathcal{M}%
}^{\left(  2\right)  }$-expansion of $\mathfrak{osp}\left(  4|1\right)  $
corresponds to the $AdS-$Lorentz superalgebra $sAdS-\mathcal{L}_{4}$ in four
dimensions. \ The details of its construction can be found in ref.
\cite{FISV}. \ An extensive study of the relation between Lie algebras and the
semigroup expansion method can be found in ref. \cite{AMNT}.

One can see that the $AdS-$Lorentz superalgebra contains the $AdS-\mathcal{L}%
_{4}$ algebra $=\left\{  J_{ab},P_{a},Z_{ab}\right\}  $%
\footnote{Also known as Poincar\'{e}
semi-simple extended
algebra.}
as a subalgebra. \ The $AdS-\mathcal{L}_{4}$ algebra and its generalization
have been extensively studied in ref. \cite{SS}. \ In particular it was shown
that this algebra allows to include a generalized cosmological constant in a
Born-Infeld gravity action.

On the other hand it is well known that an In\"{o}n\"{u}-Wigner contraction of
the $AdS-$Lorentz superalgebra leads to the Maxwell superalgebra. \ In fact,
the rescaling%
\begin{equation}
Z_{ab}\rightarrow\mu^{2}Z_{ab},\text{ \ \ }P_{a}\rightarrow\mu P_{a}\text{
\ and \ }Q_{\alpha}\rightarrow\mu Q_{\alpha}%
\end{equation}
provide the Maxwell superalgebra in the limit $\mu\rightarrow\infty$.

\section{Generalized supersymmetric cosmological term from $AdS-$Lorentz
superalgebra}

\qquad In ref. \cite{MM} it was introduced a geometric formulation of $N=1$,
$D=4$ supergravity using the $\mathfrak{osp}\left(  4|1\right)  $ gauge
fields. \ The resulting action is known as the MacDowell-Mansouri action whose
geometrical interpretation can be found in ref. \cite{DKW}. \ In a very
similar way to ref. \cite{CR2} in which a MacDowell-Mansouri like action was
built for the minimal Maxwell superalgebra, we will construct an action for
the $AdS$-Lorentz superalgebra using the useful properties of the
$S$-expansion procedure.

We have shown in the previous section that the $D=4$ $AdS-$Lorentz
superalgebra can be found as an $S$-expansion of the $\mathfrak{osp}\left(
4|1\right)  $ superalgebra. \ Following the definitions of ref. \cite{Sexp},
let $S_{\mathcal{M}}^{\left(  2\right)  }=\left\{  \lambda_{0},\lambda
_{1},\lambda_{2}\right\}  $ be an abelian semigroup whose elements satisfy the
multiplication law $\left(  \ref{lm1}\right)  .$ \ After the extraction of a
resonant subalgebra one finds the $AdS-$Lorentz superalgebra whose generators
$\left\{  J_{ab},P_{a},Z_{ab},Q_{\alpha}\right\}  $ satisfy the commutations
relations $\left(  \ref{ADSL401}\right)  -\left(  \ref{ADSL408}\right)  $.

In order to write down an action for $AdS-$Lorentz superalgebra we start from
the one-form connection%
\begin{equation}
A=A^{A}T_{A}=\frac{1}{2}\omega^{ab}J_{ab}+\frac{1}{l}e^{a}P_{a}+\frac{1}%
{2}k^{ab}Z_{ab}+\frac{1}{\sqrt{l}}\psi^{\alpha}Q_{\alpha},
\end{equation}
where the one-form gauge fields are given in terms of the components of the
$\mathfrak{osp}\left(  4|1\right)  $ connection,%
\begin{align*}
\omega^{ab}  &  =\lambda_{0}\tilde{\omega}^{ab},\\
e^{a}  &  =\lambda_{2}\tilde{e}^{a},\\
k^{ab}  &  =\lambda_{2}\tilde{\omega}^{ab},\\
\psi^{\alpha}  &  =\lambda_{1}\tilde{\psi}^{\alpha}.
\end{align*}

The associated two-form curvature $F=dA+A\wedge A$ is given by%
\begin{equation}
F=F^{A}T_{A}=\frac{1}{2}R^{ab}J_{ab}+\frac{1}{l}R^{a}P_{a}+\frac{1}{2}%
F^{ab}Z_{ab}+\frac{1}{\sqrt{l}}\Psi^{\alpha}Q_{\alpha}, \label{2curv}%
\end{equation}
where%
\begin{align*}
R^{ab}  &  =d\omega^{ab}+\omega_{\text{ }c}^{a}\omega^{cb},\\
R^{a}  &  =de^{a}+\omega_{\text{ }b}^{a}e^{b}+k_{\text{ }b}^{a}e^{b}-\frac
{1}{2}\bar{\psi}\gamma^{a}\psi,\\
F^{ab}  &  =dk^{ab}+\omega_{\text{ }c}^{a}k^{cb}-\omega_{\text{ }c}^{b}%
k^{ca}+k_{\text{ }c}^{a}k^{cb}+\frac{1}{l^{2}}e^{a}e^{b}+\frac{1}{2l}\bar
{\psi}\gamma^{ab}\psi,\\
\Psi &  =d\psi+\frac{1}{4}\omega_{ab}\gamma^{ab}\psi+\frac{1}{2l}e^{a}%
\gamma_{a}\psi+\frac{1}{4}k_{ab}\gamma^{ab}\psi.
\end{align*}
The one-forms $\omega^{ab},e^{a},\psi$ and $k^{ab}$ are the spin connection,
the vielbein, the gravitino field and a bosonic "matter" field, respectively.
\ Here $\psi$ corresponds to a Majorana spinor which satisfies $\bar{\psi
}=\psi C$, where $C$ is the charge conjugation matrix. \ Naturally when $F=0$
the Maurer-Cartan equations for the $AdS-$Lorentz superalgebra are satisfied.

In order to interpret the gauge field as the vielbein, it is necessary to
introduce a length scale $l$. \ In fact, if we choose the Lie algebra
generators $T_{A}$ to be dimensionless then the $1$-form connection fields
$A=A_{\text{ }\mu}^{A}T_{A}dx^{\mu}$ must also be dimensionless.
\ Nevertheless, the vielbein $e^{a}=e_{\text{ }\mu}^{a}dx^{\mu}$ must have
dimensions of length if it is related to the spacetime metric $g_{\mu\nu}$
through $g_{\mu\nu}=e_{\text{ }\mu}^{a}e_{\text{ }\nu}^{b}\eta_{ab}$. \ Thus
the "true" gauge field must be of the form $e^{a}/l$. \ In the same way we
must consider that $\psi/\sqrt{l}$ is the "true" gauge field of supersymmetry
since the gravitino $\psi=\psi_{\mu}dx^{\mu}$ has dimensions of $\left(
\text{length}\right)  ^{1/2}$.

From the Bianchi identity $\nabla F=0$, with $\nabla=d+\left[  A,\cdot\right]
$, it is possible to write down the Lorentz covariant exterior derivatives of
the curvatures as%
\begin{align}
DR^{ab}  &  =0,\\
DR^{a}  &  =R_{\text{ }b}^{a}e^{b}+R^{c}k_{c}^{\text{ }a}+\bar{\psi}\gamma
^{a}\Psi,\\
DF^{ab}  &  =R_{\text{ }c}^{a}k^{cb}-R_{\text{ }c}^{b}k^{ca}+F_{\text{ }c}%
^{a}k^{cb}-F_{\text{ }c}^{b}k^{ca}+\frac{1}{l^{2}}\left(  R^{a}e^{b}%
-e^{a}R^{b}\right) \nonumber\\
&  +\frac{1}{l}\bar{\Psi}\gamma^{ab}\psi,\\
D\Psi &  =\frac{1}{4}R_{ab}\gamma^{ab}\psi+\frac{1}{4}F_{ab}\gamma^{ab}%
\psi-\frac{1}{4}k_{ab}\gamma^{ab}\Psi+\frac{1}{2l}R^{a}\gamma_{a}%
\psi\nonumber\\
&  -\frac{1}{2l}e^{a}\gamma_{a}\Psi.
\end{align}

The general form of the MacDowell-Mansouri action built with$\ $the
$\mathfrak{osp}\left(  4|1\right)  $ two-form curvature is given by%
\begin{equation}
S=2\int\left\langle F\wedge F\right\rangle =2\int F^{A}\wedge F^{B}%
\left\langle T_{A}T_{B}\right\rangle ,
\end{equation}
with the following choice of the invariant tensor%
\begin{equation}
\left\langle T_{A}T_{B}\right\rangle =\left\{
\begin{array}
[c]{c}%
\left\langle J_{ab}J_{cd}\right\rangle =\epsilon_{abcd}\text{ \ \ \ \ }\\
\left\langle Q_{\alpha}Q_{\beta}\right\rangle =2\left(  \gamma_{5}\right)
_{\alpha\beta}.
\end{array}
\right.  \label{invt1}%
\end{equation}
It is important to note that if $\left\langle T_{A}T_{B}\right\rangle $ is an
invariant tensor for the $\mathfrak{osp}\left(  4|1\right)  $ superalgebra
then the action corresponds to a topological invariant. \ The action can be
seen as the supersymmetric generalization of the $D=4$ Born-Infeld action in
which the action is built from the $AdS$ two-form curvature using
$\left\langle T_{A}T_{B}\right\rangle $ as an invariant tensor for the Lorentz group.

In order to build a MacDowell-Mansouri like action for the $AdS-$Lorentz
superalgebra we will consider the $S$-expansion of $\left\langle T_{A}%
T_{B}\right\rangle $ and the $2$-form curvature given by $\left(
\ref{2curv}\right)  $.

Thus, the action for the $AdS-$Lorentz superalgebra can be written as%
\begin{equation}
S=2\int F^{A}\wedge F^{B}\left\langle T_{A}T_{B}\right\rangle
_{sAdS-\mathcal{L}_{4}},
\end{equation}
where $\left\langle T_{A}T_{B}\right\rangle _{sAdS-\mathcal{L}_{4}}$ can be
derived from the original components of the invariant tensor $\left(
\ref{invt1}\right)  $. \ Using Theorem VII.1 of ref. \cite{Sexp}, it is
possible to show that the non-vanishing components of $\left\langle T_{A}%
T_{B}\right\rangle _{sAdS-\mathcal{L}_{4}}$ are given by%
\begin{align}
\left\langle J_{ab}J_{cd}\right\rangle _{sAdS-\mathcal{L}_{4}}  &  =\alpha
_{0}\left\langle J_{ab}J_{cd}\right\rangle ,\label{inv01}\\
\left\langle J_{ab}Z_{cd}\right\rangle _{sAdS-\mathcal{L}_{4}}  &  =\alpha
_{2}\left\langle J_{ab}J_{cd}\right\rangle ,\\
\left\langle Z_{ab}Z_{cd}\right\rangle _{sAdS-\mathcal{L}_{4}}  &  =\alpha
_{2}\left\langle J_{ab}J_{cd}\right\rangle ,\\
\left\langle Q_{\alpha}Q_{\beta}\right\rangle _{sAdS-\mathcal{L}_{4}}  &
=\alpha_{2}\left\langle Q_{\alpha}Q_{\beta}\right\rangle , \label{inv04}%
\end{align}
where $\alpha_{0}$ and $\alpha_{2}$ are dimensionless arbitrary independent
constants. \ This choice of the invariant tensor breaks the $AdS$-Lorentz
supergroup to their Lorentz like subgroup.

Then considering the non-vanishing components of the invariant tensor $\left(
\ref{inv01}\right)  -\left(  \ref{inv04}\right)  $ and the $2$-form curvature
$\left(  \ref{2curv}\right)  $, it is possible to write down an action as%
\begin{equation}
S=2\int\left(  \frac{1}{4}\alpha_{0}\epsilon_{abcd}R^{ab}R^{cd}+\frac{1}%
{2}\alpha_{2}\epsilon_{abcd}R^{ab}F^{cd}+\frac{1}{4}\alpha_{2}\epsilon
_{abcd}F^{ab}F^{cd}+\frac{2}{l}\alpha_{2}\bar{\Psi}\gamma_{5}\Psi\right)  .
\end{equation}
Explicitly, the action takes the form%
\begin{align}
S &  =\int\frac{\alpha_{0}}{2}\epsilon_{abcd}R^{ab}R^{cd}+\alpha_{2}%
\epsilon_{abcd}\left(  R^{ab}Dk^{cd}+R^{ab}k_{\text{ }e}^{c}k^{ed}+\frac
{1}{l^{2}}R^{ab}e^{c}e^{d}\right.  \nonumber\\
&  \left.  +\frac{1}{2l}R^{ab}\bar{\psi}\gamma^{cd}\psi+\frac{1}{2}%
Dk^{ab}Dk^{cd}+Dk^{ab}k_{\text{ }e}^{c}k^{ed}+\frac{1}{l^{2}}Dk^{ab}e^{c}%
e^{d}\right.  \nonumber\\
&  \left.  +\frac{1}{2l}Dk^{ab}\bar{\psi}\gamma^{cd}\psi+\frac{1}{2}k_{\text{
}f}^{a}k^{fb}k_{\text{ }g}^{c}k^{gd}+\frac{1}{l^{2}}k_{\text{ }f}^{a}%
k^{fb}e^{c}e^{d}+\frac{1}{2l}k_{\text{ }f}^{a}k^{fb}\bar{\psi}\gamma^{cd}%
\psi\right.  \nonumber\\
&  \left.  +\frac{1}{2l^{3}}e^{a}e^{b}\bar{\psi}\gamma^{cd}\psi+\frac
{1}{2l^{4}}e^{a}e^{b}e^{c}e^{d}\right)  +\alpha_{2}\left(  \frac{4}{l}%
D\bar{\psi}\gamma_{5}D\psi+\frac{4}{l^{2}}\bar{\psi}e^{a}\gamma_{a}\gamma
_{5}D\psi\right.  \nonumber\\
&  \left.  +\frac{2}{l}D\bar{\psi}\gamma_{5}k_{ab}\gamma^{ab}\psi+\frac
{1}{l^{3}}\bar{\psi}e^{a}\gamma_{a}\gamma_{5}e^{b}\gamma_{b}\psi+\frac
{1}{l^{2}}\bar{\psi}e^{a}\gamma_{a}\gamma_{5}k^{bc}\gamma_{bc}\psi\right.
\nonumber\\
&  \left.  +\frac{1}{4l}\bar{\psi}k_{ab}\gamma^{ab}\gamma_{5}k_{cd}\gamma
^{cd}\psi\right)  .
\end{align}
The action can be written in a more compact way using the gamma matrix
identity%
\begin{equation}
\gamma_{ab}\gamma_{5}=-\frac{1}{2}\epsilon_{abcd}\gamma^{cd},
\end{equation}
and the gravitino Bianchi identity%
\begin{equation}
DD\psi=\frac{1}{4}R^{ab}\gamma_{ab}\psi.
\end{equation}
In fact one can see that%
\begin{align*}
\frac{1}{2}\epsilon_{abcd}R^{ab}\bar{\psi}\gamma^{cd}\psi+4D\bar{\psi}%
\gamma_{5}D\psi &  =d\left(  4D\bar{\psi}\gamma_{5}\psi\right)  ,\\
\frac{1}{2}\epsilon_{abcd}Dk^{ab}\bar{\psi}\gamma^{cd}\psi+2D\bar{\psi}%
\gamma_{5}k^{ab}\gamma_{ab}\psi &  =d\left(  \bar{\psi}k^{ab}\gamma_{ab}%
\gamma_{5}\psi\right)  .
\end{align*}
Furthermore it is possible to show that%
\begin{align*}
\bar{\psi}e^{a}\gamma_{a}\gamma_{5}e^{b}\gamma_{b}\psi &  =\frac{1}{2}%
e^{a}e^{b}\bar{\psi}\gamma^{cd}\psi\epsilon_{abcd},\\
\frac{1}{4}\bar{\psi}k_{ab}\gamma^{ab}\gamma_{5}k_{cd}\gamma^{cd}\psi &
=-\frac{1}{2}k_{\text{ }f}^{a}k^{fb}\bar{\psi}\gamma^{cd}\psi\epsilon
_{abcd},\\
\bar{\psi}e^{a}\gamma_{a}\gamma_{5}k^{bc}\gamma_{bc}\psi &  =\epsilon
_{abcd}k^{ab}e^{c}\bar{\psi}\gamma^{d}\psi,
\end{align*}
where we have used the following identities%
\begin{align*}
\gamma_{a}\gamma_{b} &  =\gamma_{ab}+\eta_{ab},\\
\gamma^{ab}\gamma^{cd} &  =\epsilon^{abcd}\gamma_{5}-4\delta_{\lbrack c}%
^{[a}\gamma_{d]}^{b]}-2\delta_{cd}^{ab},\\
\gamma^{c}\gamma^{ab} &  =-2\gamma^{\lbrack a}\delta_{c}^{b]}-\epsilon
^{abcd}\gamma_{5}\gamma_{d},
\end{align*}
and the fact that $\gamma_{5}\gamma_{a}$ is an antisymmetric matrix. \ Thus
the MacDowell-Mansouri like action for the $AdS-$Lorentz superalgebra takes
the form%
\begin{align}
S &  =\int\frac{\alpha_{0}}{2}\epsilon_{abcd}R^{ab}R^{cd}+\frac{\alpha_{2}%
}{l^{2}}\left(  \epsilon_{abcd}R^{ab}e^{c}e^{d}+4\bar{\psi}e^{a}\gamma
_{a}\gamma_{5}D\psi\right)  \nonumber\\
&  +\alpha_{2}\epsilon_{abcd}\left(  R^{ab}Dk^{cd}+R^{ab}k_{\text{ }e}%
^{c}k^{ed}+\frac{1}{2}Dk^{ab}Dk^{cd}+Dk^{ab}k_{\text{ }e}^{c}k^{ed}+\frac
{1}{2}k_{\text{ }f}^{a}k^{fb}k_{\text{ }g}^{c}k^{gd}\right)  \nonumber\\
&  +\alpha_{2}\epsilon_{abcd}\left(  \frac{1}{l^{2}}Dk^{ab}e^{c}e^{d}+\frac
{1}{l^{2}}k_{\text{ }f}^{a}k^{fb}e^{c}e^{d}+\frac{1}{l^{3}}e^{a}e^{b}\bar
{\psi}\gamma^{cd}\psi\right.  \nonumber\\
&  \left.  +\frac{1}{l^{2}}k^{ab}e^{c}\bar{\psi}\gamma^{d}\psi+\frac{1}%
{2l^{4}}e^{a}e^{b}e^{c}e^{d}\right)  +\alpha_{2}d\left(  4D\bar{\psi}%
\gamma_{5}\psi+\bar{\psi}k^{ab}\gamma_{ab}\gamma_{5}\psi\right)
.\label{actionsAdSL4}%
\end{align}

This action has been intentionally separated in five pieces where the first
term is proportional to $\alpha_{0}$ and corresponds to the Gauss Bonnet term.
\ The second term contains the Einstein-Hilbert term plus the Rarita-Schwinger
(RS) Lagrangian describing pure supergravity.\ The third piece corresponds to
a Gauss Bonnet like term containing the new super $AdS-$Lorentz fields. \ This
piece does not contribute to the dynamics and can be written as a boundary
term. \ The fourth term corresponds to a generalized supersymmetric
cosmological term which contains the usual supersymmetric cosmological
constant plus three additional terms depending on $k^{ab}$. \ The last piece
is a boundary term.

One can see that the MacDowell-Mansouri like action built using the useful
definitions of the $S$-expansion procedure describes a supergravity theory
with a generalized supersymmetric cosmological term.

From $\left(  \ref{actionsAdSL4}\right)  $ we can see that the bosonic part of
the action corresponds to the one found for $AdS-$Lorentz algebra in ref.
\cite{SS}. \ Besides, the action contains the generalized cosmological term
introduced in ref. \cite{AKL} for the Maxwell algebra.

One can note that if we omit the boundary terms in $\left(  \ref{actionsAdSL4}%
\right)  $, the action can be written as
\begin{align}
S  &  =\int\frac{\alpha_{2}}{l^{2}}\left(  \epsilon_{abcd}R^{ab}e^{c}%
e^{d}+4\bar{\psi}e^{a}\gamma_{a}\gamma_{5}D\psi\right)  +\alpha_{2}%
\epsilon_{abcd}\left(  \frac{1}{l^{2}}Dk^{ab}e^{c}e^{d}+\frac{1}{l^{2}%
}k_{\text{ }f}^{a}k^{fb}e^{c}e^{d}\right. \nonumber\\
&  \left.  +\frac{1}{l^{3}}e^{a}e^{b}\bar{\psi}\gamma^{cd}\psi+\frac{1}{l^{2}%
}k^{ab}e^{c}\bar{\psi}\gamma^{d}\psi+\frac{1}{2l^{4}}e^{a}e^{b}e^{c}%
e^{d}\right)  ,
\end{align}
or equivalently%
\begin{align}
S  &  =\int\frac{\alpha_{2}}{l^{2}}\left(  \epsilon_{abcd}R^{ab}e^{c}%
e^{d}+4\bar{\psi}e^{a}\gamma_{a}\gamma_{5}D\psi\right) \nonumber\\
&  +\alpha_{2}\epsilon_{abcd}\left(  \frac{2}{l^{2}}k^{ab}\hat{T}^{c}%
e^{d}+\frac{1}{l^{2}}k_{\text{ }f}^{a}k^{fb}e^{c}e^{d}+\frac{1}{l^{3}}%
e^{a}e^{b}\bar{\psi}\gamma^{cd}\psi+\frac{1}{2l^{4}}e^{a}e^{b}e^{c}%
e^{d}\right)  ,
\end{align}
where we have used%
\begin{align*}
{\Large \epsilon}_{abcd}{\Large Dk}^{ab}{\Large e}^{c}{\Large e}^{d}  &
=2{\Large \epsilon}_{abcd}{\Large k}^{ab}{\Large T}^{c}{\Large e}^{d}+d\left(
\frac{1}{l^{2}}{\Large \epsilon}_{abcd}{\Large k}^{ab}{\Large e}^{c}%
{\Large e}^{d}\right)  ,\\
\hat{T}^{a}  &  =De^{a}-\frac{1}{2}\bar{\psi}\gamma^{a}\psi=T^{a}-\frac{1}%
{2}\bar{\psi}\gamma^{a}\psi.
\end{align*}
Interestingly if we consider $k^{ab}=0$ in our action we obtain the usual
MacDowell-Mansouri action for the $Osp\left(  4|1\right)  $ supergroup.

In order to obtain the field equations let us compute the variation of the
Lagrangian with respect to the different super $AdS-$Lorentz fields. \ The
variation of the Lagrangian with respect to the spin connection $\omega^{ab}$,
modulo boundary terms, is given by%
\begin{align}
\delta_{\omega}\mathcal{L}  &  =\frac{\alpha_{2}}{l^{2}}\epsilon_{abcd}\left(
2\delta\omega^{ab}De^{c}e^{d}+2\delta\omega_{\text{ }f}^{a}k^{fb}e^{c}%
e^{d}\right)  +\frac{\alpha_{2}}{l^{2}}\bar{\psi}e^{a}\gamma_{a}\gamma
_{5}\delta\omega^{cd}\gamma_{cd}\psi\nonumber\\
&  =\frac{2\alpha_{2}}{l^{2}}\epsilon_{abcd}\delta\omega^{ab}\left(
T^{c}+k_{\text{ }f}^{c}e^{f}-\frac{1}{2}\bar{\psi}\gamma^{c}\psi\right)
e^{d}\nonumber\\
&  =\frac{2\alpha_{2}}{l^{2}}\epsilon_{abcd}\delta\omega^{ab}R^{c}e^{d}.
\end{align}
Here we see that $\delta_{\omega}\mathcal{L}=0$ leads to the following field
equation for the $AdS-$Lorentz supertorsion%
\begin{equation}
\epsilon_{abcd}R^{a}e^{d}=0. \label{eqomega}%
\end{equation}

On the other hand, the variation of the Lagrangian with respect to the
vielbein $e^{a}$ is given by%
\begin{align}
\delta_{e}\mathcal{L}  &  =\frac{\alpha_{2}}{l^{2}}\epsilon_{abcd}\left(
2R^{ab}e^{c}+2Dk^{ab}e^{c}+2k_{\text{ }f}^{a}k^{fb}e^{c}+\frac{2}{l}\bar{\psi
}\gamma^{ab}\psi e^{c}+\frac{2}{l^{2}}e^{a}e^{b}e^{c}\right)  \delta
e^{d}\nonumber\\
&  +\frac{\alpha_{2}}{l^{2}}\left(  4\bar{\psi}\gamma_{d}\gamma_{5}D\psi
+\bar{\psi}\gamma_{d}\gamma_{5}k^{ab}\gamma_{ab}\psi\right)  \delta
e^{d}.\nonumber\\
&  =\frac{2\alpha_{2}}{l^{2}}\epsilon_{abcd}\left(  R^{ab}e^{c}+F^{ab}%
e^{c}\right)  \delta e^{d}+\frac{\alpha_{2}}{l^{2}}\left(  4\bar{\psi}%
\gamma_{d}\gamma_{5}\Psi\right)  \delta e^{d}, \label{vare3}%
\end{align}
where we have used the $AdS-$Lorentz $2$-form curvatures $\left(
\ref{2curv}\right)  $ and the fact that
\begin{align*}
\epsilon_{abcd}\bar{\psi}\gamma^{ab}\psi e^{c}  &  =2\bar{\psi}\gamma
_{d}\gamma_{5}e^{c}\gamma_{c}\psi,\\
\epsilon_{abcd}k^{ab}e^{c}\bar{\psi}\gamma^{d}\psi &  =\bar{\psi}e^{a}%
\gamma_{a}\gamma_{5}k^{bc}\gamma_{bc}\psi.
\end{align*}
Then the field equation is obtained imposing $\delta_{e}\mathcal{L}=0$%
\begin{equation}
2\epsilon_{abcd}\left(  R^{ab}+F^{ab}\right)  e^{c}+4\bar{\psi}\gamma
_{d}\gamma_{5}\Psi=0.
\end{equation}
One can see that the rescaling%
\[
k^{ab}\rightarrow\mu^{2}k^{ab}\text{, \ \ }e^{a}\rightarrow\mu e^{a}\text{
\ and\ \ }\psi\rightarrow\sqrt{\mu}\psi
\]
and dividing $\left(  \ref{vare3}\right)  $ by $\mu^{2}$ provide us with the
usual field equation for supergravity in the limit $\mu\rightarrow0,$%
\begin{equation}
\epsilon_{abcd}R^{ab}e^{c}+4\bar{\psi}\gamma_{d}\gamma_{5}D\psi=0,
\end{equation}
where $D$ corresponds to the Lorentz covariant exterior derivative.

The variation of the Lagrangian with respect to the new $AdS-$Lorentz field
$k^{ab}$, modulo boundary terms, gives%
\begin{align}
\delta_{k}\mathcal{L}  &  =\frac{\alpha_{2}}{l^{2}}\epsilon_{abcd}\left(
2\delta k^{ab}De^{c}e^{d}+2\delta k_{\text{ }f}^{a}k^{fb}e^{c}e^{d}+\frac
{1}{l^{2}}\delta k^{ab}\bar{\psi}\gamma^{d}\psi e^{c}\right) \nonumber\\
&  =\frac{2\alpha_{2}}{l^{2}}\epsilon_{abcd}\delta k^{ab}\left(
T^{c}+k_{\text{ }f}^{c}e^{f}-\frac{1}{2}\bar{\psi}\gamma^{c}\psi\right)
e^{d}\nonumber\\
&  =\frac{2\alpha_{2}}{l^{2}}\epsilon_{abcd}\delta k^{ab}R^{c}e^{d},
\end{align}
where we have used the gamma matrix identities%
\begin{align*}
\gamma_{ab}\gamma_{5}  &  =-\frac{1}{2}\epsilon_{abcd}\gamma^{cd},\\
\gamma^{c}\gamma^{ab}  &  =-2\gamma^{\lbrack a}\delta_{c}^{b]}-\epsilon
^{abcd}\gamma_{5}\gamma_{d}.
\end{align*}
Here we see that $\delta_{k}\mathcal{L}=0$ leads to the same field equation
than $\delta_{\omega}\mathcal{L}=0$%
\begin{equation}
\epsilon_{abcd}R^{a}e^{d}=0. \label{eqkab}%
\end{equation}

Let us consider the variation of the Lagrangian with respect to the gravitino
field $\psi$, modulo boundary terms,%
\begin{align}
\delta_{\psi}\mathcal{L}  &  =\frac{\alpha_{2}}{l^{2}}\left(  4\delta\bar
{\psi}e^{a}\gamma_{a}\gamma_{5}D\psi+4D\bar{\psi}e^{a}\gamma_{a}\gamma
_{5}\delta\psi-4\bar{\psi}De^{a}\gamma_{a}\gamma_{5}\delta\psi\right.
\nonumber\\
&  \left.  +2\delta\bar{\psi}e^{a}\gamma_{a}\gamma_{5}k^{bc}\gamma_{bc}%
\psi+4\delta\bar{\psi}\gamma_{a}\gamma_{5}k_{b}^{a}e^{b}\psi\right)
+\frac{\alpha_{2}}{l^{2}}\epsilon_{abcd}\left(  \frac{2}{l}e^{a}e^{b}%
\delta\bar{\psi}\gamma^{cd}\psi\right) \nonumber\\
&  =\frac{\alpha_{2}}{l^{2}}\left(  4\delta\bar{\psi}e^{a}\gamma_{a}\gamma
_{5}D\psi+4\delta\bar{\psi}\gamma_{a}\gamma_{5}D\psi e^{a}+4\delta\bar{\psi
}De^{a}\gamma_{a}\gamma_{5}\psi\right. \nonumber\\
&  \left.  +2\delta\bar{\psi}e^{a}\gamma_{a}\gamma_{5}k^{bc}\gamma_{bc}%
\psi+4\delta\bar{\psi}\gamma_{a}\gamma_{5}k_{b}^{a}e^{b}\psi\right)
+\frac{4\alpha_{2}}{l^{3}}\delta\bar{\psi}e^{a}\gamma_{a}\gamma_{5}e^{b}%
\gamma_{b}\psi\nonumber\\
&  =\frac{\alpha_{2}}{l^{2}}\delta\bar{\psi}\left(  8e^{a}\gamma_{a}\gamma
_{5}\Psi+4\gamma_{a}\gamma_{5}\psi De^{a}+4\gamma_{a}\gamma_{5}k_{b}^{a}%
e^{b}\psi\right)  .
\end{align}
Then, using the definition of the supertorsion%
\[
R^{a}=De^{a}+k_{\text{ }b}^{a}e^{b}-\frac{1}{2}\bar{\psi}\gamma^{a}\psi,
\]
and the Fierz identity%
\[
\gamma_{a}\psi\bar{\psi}\gamma^{a}\psi=0,
\]
we find the following field equation,%
\begin{equation}
8e^{a}\gamma_{a}\gamma_{5}\Psi+4\gamma_{a}\gamma_{5}\psi R^{a}=0.
\end{equation}
We can see that the introduction of a generalized supersymmetric cosmological
constant leads to field equations very similar to those of $\mathfrak{osp}%
\left(  4|1\right)  $ supergravity. \ The differences appear in the definition
of the two-form curvatures due to the presence of the new matter field
$k^{ab}$.

Let us note that, from eqs $\left(  \ref{eqomega}\right)  $ and $\left(
\ref{eqkab}\right)  $, the equation of motion coming from the variation of the
Lagrangian with respect to the bosonic field $k^{ab}$ reduces to that of the
spin connection $\omega^{ab}$.%
\begin{equation}
\epsilon_{abcd}R^{a}e^{d}=\epsilon_{abcd}\left(  T^{c}+k_{\text{ }f}^{c}%
e^{f}-\frac{1}{2}\bar{\psi}\gamma^{c}\psi\right)  e^{d}=0.
\end{equation}
Interestingly, we can define a new bosonic field as%
\begin{equation}
\varpi^{ab}=\omega^{ab}+k^{ab},
\end{equation}
and its respective covariant derivative,%
\begin{equation}
\mathcal{D}=d+\varpi.
\end{equation}
Then, the equation of motion can be written as%
\begin{equation}
\epsilon_{abcd}\left(  \mathcal{D}e^{c}-\frac{1}{2}\bar{\psi}\gamma^{c}%
\psi\right)  e^{d}=0.
\end{equation}
This allows to express the bosonic field $\varpi^{ab}$ in terms of the
vielbein $e^{a}$ and gravitino fields $\psi^{\alpha}$. \ This may be solved
considering the following decomposition,%
\begin{equation}
\varpi^{ab}=\mathring{\varpi}^{ab}+\tilde{\varpi}^{ab},
\end{equation}
where $\mathring{\varpi}^{ab}$ corresponds to the solution of $\mathcal{D}%
e^{c}=0$ and it is given by%
\begin{equation}
\mathring{\varpi}_{\mu}^{ab}=\left(  e_{\lambda}^{c}\partial_{\lbrack\mu
}e_{\nu]}^{d}\eta_{cd}+e_{\nu}^{c}\partial_{\lbrack\lambda}e_{\mu]}^{d}%
\eta_{cd}-e_{\mu}^{c}\partial_{\lbrack\nu}e_{\lambda]}^{d}\eta_{cd}\right)
e^{\lambda|a}e^{\nu|b}.
\end{equation}
Now we have that%
\begin{equation}
\mathcal{D}e^{a}=de^{a}+\mathring{\varpi}^{ab}e_{b}+\tilde{\varpi}^{ab}%
e_{b}=\frac{1}{2}\bar{\psi}\gamma^{a}\psi,
\end{equation}
implies%
\begin{equation}
\tilde{\varpi}_{[\mu}^{ab}e_{\nu]b}=\frac{1}{2}\bar{\psi}_{\mu}\gamma^{a}%
\psi_{\nu}.
\end{equation}
Then we may solve $\tilde{\varpi}^{ab}$ in terms of the two other fields,
\begin{equation}
\tilde{\varpi}_{\mu}^{ab}=\frac{1}{4}e^{a|\lambda}e^{b|\nu}\left(  \bar{\psi
}_{\mu}\gamma_{\lambda}\psi_{\nu}+\bar{\psi}_{\lambda}\gamma_{\nu}\psi_{\mu
}-\bar{\psi}_{\nu}\gamma_{\mu}\psi_{\lambda}-\bar{\psi}_{\mu}\gamma_{\nu}%
\psi_{\lambda}-\bar{\psi}_{\nu}\gamma_{\lambda}\psi_{\mu}+\bar{\psi}_{\lambda
}\gamma_{\mu}\psi_{\nu}\right)  .
\end{equation}
Thus, the bosonic field $\varpi^{ab}$ is completely determined in terms of
$e_{\mu}^{a}$ and $\psi_{\mu}^{\alpha}$ and does not carry additional physical
degrees of freedom. \ In particular, when the supertorsion $R^{a}%
=\mathcal{D}e^{c}-\frac{1}{2}\bar{\psi}\gamma^{c}\psi$ is set equal to zero,
the number of bosonic degrees of freedom is two as the Einstein-Hilbert
gravity theory and corresponds to the remaining components of the vielbein.

On the other hand, although the Lagrangian is built from the $AdS$-Lorentz
superalgebra it is not invariant under gauge transformations. \ In fact, the
Lagrangian does not correspond to a Yang-Mills Lagrangian, nor a topological invariant.

As we can see the variation of the action $\left(  \ref{actionsAdSL4}\right)
$ under gauge supersymmetry can be obtained using $\delta R=\left[
\epsilon,R\right]  $,%
\begin{equation}
\delta_{susy}S=-\frac{4\alpha_{2}}{l^{2}}\int R^{a}\bar{\Psi}\gamma_{a}%
\gamma_{5}\epsilon.
\end{equation}
Thus in order to have gauge supersymmetry invariance it is necessary to impose
the supertorsion constraint%
\begin{equation}
R^{a}=0\text{. }%
\end{equation}
However this leads to express the spin connection $\omega^{ab}$ in terms of
the others fields $\left\{  e^{a},k^{ab},\psi\right\}  $.

Nevertheless, it is possible to have supersymmetry invariance in the first
formalism adding an extra piece to the gauge transformation $\delta\omega
^{ab}$ such that the variation of the action can be written as%
\begin{equation}
\delta S=-\frac{4\alpha_{2}}{l^{2}}\int R^{a}\left[  \bar{\Psi}\gamma
_{a}\gamma_{5}\epsilon-\frac{1}{2}\epsilon_{abcd}e^{b}\delta_{extra}%
\omega^{cd}\right]  ,
\end{equation}
where the supersymmetry invariance is fullfilled when%
\begin{equation}
\delta_{extra}\omega^{ab}=2\epsilon^{abcd}\left(  \bar{\Psi}_{ec}\gamma
_{d}\gamma_{5}\epsilon+\bar{\Psi}_{de}\gamma_{c}\gamma_{5}\epsilon-\bar{\Psi
}_{cd}\gamma_{e}\gamma_{5}\epsilon\right)  e^{e},
\end{equation}
with $\bar{\Psi}=\bar{\Psi}_{ab}e^{a}e^{b}$.

Thus, the action $\left(  \ref{actionsAdSL4}\right)  $ in the first order
formalism is invariant under the following supersymmetry transformations%
\begin{align}
\delta\omega^{ab}  &  =2\epsilon^{abcd}\left(  \bar{\Psi}_{ec}\gamma_{d}%
\gamma_{5}\epsilon+\bar{\Psi}_{de}\gamma_{c}\gamma_{5}\epsilon-\bar{\Psi}%
_{cd}\gamma_{e}\gamma_{5}\epsilon\right)  e^{e},\\
\delta k^{ab}  &  =-\frac{1}{l}\bar{\epsilon}\gamma^{ab}\psi,\\
\delta e^{a}  &  =\bar{\epsilon}\gamma^{a}\psi,\\
\delta\psi &  =d\epsilon+\frac{1}{4}\omega^{ab}\gamma_{ab}\epsilon+\frac{1}%
{4}k^{ab}\gamma_{ab}\epsilon+\frac{1}{2l}e^{a}\gamma_{a}\epsilon.
\end{align}
Let us note that supersymmetry is not a gauge symmetry of the action, since it
is broken to a Lorentz like symmetry. \ In particular, the supersymmetry
transformations leaving the action invariant do not close off-shell. \ While
the super $AdS$-Lorentz gauge variation close off-shell by construction.

\section{The Generalized minimal $AdS-$Lorentz superalgebra}

\qquad In this section, we show that a particular choice of an abelian
semigroup $S$ leads to a new Lie superalgebra. \ For this purpose we will
consider the $\mathfrak{osp}\left(  4|1\right)  $ superalgebra as a starting point.

Let us consider a decomposition of the original superalgebra $\mathfrak{g}%
=\mathfrak{osp}\left(  4|1\right)  $ as%
\begin{align}
\mathfrak{g}=\mathfrak{osp}\left(  4|1\right)   &  =\mathfrak{so}\left(
3,1\right)  \oplus\frac{\mathfrak{osp}\left(  4|1\right)  }{\mathfrak{sp}%
\left(  4\right)  }\oplus\frac{\mathfrak{sp}\left(  4\right)  }{\mathfrak{so}%
\left(  3,1\right)  }\nonumber\\
&  =V_{0}\oplus V_{1}\oplus V_{2},
\end{align}
where $V_{0}$, $V_{1}$ and $V_{2}$ satisfy $\left(  \ref{subdes1}\right)  $
and correspond to the Lorentz subspace, the fermionic subspace and the
$AdS$-boost, respectively.

Let $S_{\mathcal{M}}^{\left(  4\right)  }=\left\{  \lambda_{0},\lambda
_{1},\lambda_{2},\lambda_{3},\lambda_{4}\right\}  $ be the abelian semigroup
whose elements satisfy the following multiplication law%
\begin{equation}
\lambda_{\alpha}\lambda_{\beta}=\left\{
\begin{array}
[c]{c}%
\lambda_{\alpha+\beta},\text{ \ \ if }\alpha+\beta\leq4\\
\lambda_{\alpha+\beta-4},\text{ \ if }\alpha+\beta>4\text{\ }%
\end{array}
\right.  \label{Slaw2}%
\end{equation}

Let us consider the decomposition $S=S_{0}\cup S_{1}\cup S_{2}$, with%
\begin{align}
S_{0}  &  =\left\{  \lambda_{0},\lambda_{2},\lambda_{4}\right\}  ,\\
S_{1}  &  =\left\{  \lambda_{1},\lambda_{3}\right\}  ,\\
S_{2}  &  =\left\{  \lambda_{2},\lambda_{4}\right\}  .
\end{align}
One can see that this decomposition satisfies the same structure as the
subspaces $V_{p}$, then we say that the decomposition is resonant [compare
with eqs $\left(  \ref{subdes1}\right)  $]%
\begin{equation}%
\begin{tabular}
[c]{ll}%
$S_{0}\cdot S_{0}\subset S_{0},$ & $S_{1}\cdot S_{1}\subset S_{0}\cap S_{2}%
,$\\
$S_{0}\cdot S_{1}\subset S_{1},$ & $S_{1}\cdot S_{2}\subset S_{1},$\\
$S_{0}\cdot S_{2}\subset S_{2},$ & $S_{2}\cdot S_{2}\subset S_{0}.$%
\end{tabular}
\end{equation}
Following theorem IV.2 of ref. \cite{Sexp}, we say that the superalgebra%
\[
\mathfrak{G}_{R}=W_{0}\oplus W_{1}\oplus W_{2}%
\]
is a resonant subalgebra of $S_{\mathcal{M}}^{\left(  4\right)  }%
\times\mathfrak{g}$, where%
\begin{align}
W_{0}  &  =\left(  S_{0}\times V_{0}\right)  =\left\{  \lambda_{0},\lambda
_{2},\lambda_{4}\right\}  \times\left\{  \tilde{J}_{ab}\right\}  =\left\{
\lambda_{0}\tilde{J}_{ab},\lambda_{2}\tilde{J}_{ab},\lambda_{4}\tilde{J}%
_{ab}\right\}  ,\\
W_{1}  &  =\left(  S_{1}\times V_{1}\right)  =\left\{  \lambda_{1},\lambda
_{3}\right\}  \times\left\{  \tilde{Q}_{\alpha}\right\}  =\left\{  \lambda
_{1}\tilde{Q}_{\alpha},\lambda_{3}\tilde{Q}_{\alpha}\right\}  ,\\
W_{2}  &  =\left(  S_{2}\times V_{2}\right)  =\left\{  \lambda_{2},\lambda
_{4}\right\}  \times\left\{  \tilde{P}_{a}\right\}  =\left\{  \lambda
_{2}\tilde{P}_{a},\lambda_{4}\tilde{P}_{a}\right\}  .
\end{align}
Then the new superalgebra is generated by $\left\{  J_{ab},P_{a},\tilde{Z}%
_{a},\tilde{Z}_{ab},Z_{ab},Q_{\alpha},\Sigma_{\alpha}\right\}  $ with%
\begin{equation}%
\begin{tabular}
[c]{ll}%
$J_{ab}=\lambda_{0}\tilde{J}_{ab},$ & $P_{a}=\lambda_{2}\tilde{P}_{a},$\\
$\tilde{Z}_{ab}=\lambda_{2}\tilde{J}_{ab},$ & $\tilde{Z}_{a}=\lambda_{4}%
\tilde{P}_{a},$\\
$Z_{ab}=\lambda_{4}\tilde{J}_{ab},$ & $Q_{\alpha}=\lambda_{1}\tilde{Q}%
_{\alpha},$\\
$\Sigma_{\alpha}=\lambda_{3}\tilde{Q}_{\alpha}.$ &
\end{tabular}
\ \ \ \ \ \ \ \ \
\end{equation}
where $\tilde{J}_{ab},\tilde{P}_{a}$ and $\tilde{Q}_{\alpha}$ are the
$\mathfrak{osp}\left(  4|1\right)  $ generators. \ The new generators satisfy
the (anti)commutation relations%
\begin{align}
\left[  J_{ab},J_{cd}\right]   &  =\eta_{bc}J_{ad}-\eta_{ac}J_{bd}-\eta
_{bd}J_{ac}+\eta_{ad}J_{bc},\\
\left[  Z_{ab},Z_{cd}\right]   &  =\eta_{bc}Z_{ad}-\eta_{ac}Z_{bd}-\eta
_{bd}Z_{ac}+\eta_{ad}Z_{bc},\\
\left[  J_{ab},Z_{cd}\right]   &  =\eta_{bc}Z_{ad}-\eta_{ac}Z_{bd}-\eta
_{bd}Z_{ac}+\eta_{ad}Z_{bc},\\
\left[  J_{ab},\tilde{Z}_{cd}\right]   &  =\eta_{bc}\tilde{Z}_{ad}-\eta
_{ac}\tilde{Z}_{bd}-\eta_{bd}\tilde{Z}_{ac}+\eta_{ad}\tilde{Z}_{bc},\\
\left[  \tilde{Z}_{ab},\tilde{Z}_{cd}\right]   &  =\eta_{bc}Z_{ad}-\eta
_{ac}Z_{bd}-\eta_{bd}Z_{ac}+\eta_{ad}Z_{bc},\\
\left[  \tilde{Z}_{ab},Z_{cd}\right]   &  =\eta_{bc}\tilde{Z}_{ad}-\eta
_{ac}\tilde{Z}_{bd}-\eta_{bd}\tilde{Z}_{ac}+\eta_{ad}\tilde{Z}_{bc},\\
\left[  J_{ab},P_{c}\right]   &  =\eta_{bc}P_{a}-\eta_{ac}P_{b},\text{
\ \ \ \ }\left[  Z_{ab},P_{c}\right]  =\eta_{bc}P_{a}-\eta_{ac}P_{b},\text{
}\\
\left[  \tilde{Z}_{ab},P_{c}\right]   &  =\eta_{bc}\tilde{Z}_{a}-\eta
_{ac}\tilde{Z}_{b},\text{ \ \ \ \ }\left[  J_{ab},\tilde{Z}_{c}\right]
=\eta_{bc}\tilde{Z}_{a}-\eta_{ac}\tilde{Z}_{b},\\
\left[  \tilde{Z}_{ab},\tilde{Z}_{c}\right]   &  =\eta_{bc}P_{a}-\eta
_{ac}P_{b},\text{ \ \ \ \ }\left[  Z_{ab},\tilde{Z}_{c}\right]  =\eta
_{bc}\tilde{Z}_{a}-\eta_{ac}\tilde{Z}_{b},\label{dif1}\\
\left[  P_{a},P_{b}\right]   &  =Z_{ab},\text{ \ \ \ \ }\left[  \tilde{Z}%
_{a},P_{b}\right]  =\tilde{Z}_{ab},\text{ \ \ \ \ }\left[  \tilde{Z}%
_{a},\tilde{Z}_{b}\right]  =Z_{ab}, \label{dif2}%
\end{align}%
\begin{align}
\left[  J_{ab},Q_{\alpha}\right]   &  =-\frac{1}{2}\left(  \gamma
_{ab}Q\right)  _{\alpha},\text{ \ \ \ \ }\left[  P_{a},Q_{\alpha}\right]
=-\frac{1}{2}\left(  \gamma_{a}\Sigma\right)  _{\alpha},\\
\left[  \tilde{Z}_{ab},Q_{\alpha}\right]   &  =-\frac{1}{2}\left(  \gamma
_{ab}\Sigma\right)  _{\alpha},\text{ \ \ \ \ }\left[  \tilde{Z}_{a},Q_{\alpha
}\right]  =-\frac{1}{2}\left(  \gamma_{a}Q\right)  _{\alpha},\\
\left[  Z_{ab},Q_{\alpha}\right]   &  =-\frac{1}{2}\left(  \gamma
_{ab}Q\right)  _{\alpha},\text{ \ \ \ \ }\left[  P_{a},\Sigma_{\alpha}\right]
=-\frac{1}{2}\left(  \gamma_{a}Q\right)  _{\alpha},\\
\left[  J_{ab},\Sigma_{\alpha}\right]   &  =-\frac{1}{2}\left(  \gamma
_{ab}\Sigma\right)  _{\alpha},\text{ \ \ \ \ }\left[  \tilde{Z}_{a}%
,\Sigma_{\alpha}\right]  =-\frac{1}{2}\left(  \gamma_{a}\Sigma\right)
_{\alpha},\\
\left[  \tilde{Z}_{ab},\Sigma_{\alpha}\right]   &  =-\frac{1}{2}\left(
\gamma_{ab}Q\right)  _{\alpha},\text{ \ \ \ \ }\left[  Z_{ab},\Sigma_{\alpha
}\right]  =-\frac{1}{2}\left(  \gamma_{ab}\Sigma\right)  _{\alpha},\\
\left\{  Q_{\alpha},Q_{\beta}\right\}   &  =-\frac{1}{2}\left[  \left(
\gamma^{ab}C\right)  _{\alpha\beta}\tilde{Z}_{ab}-2\left(  \gamma^{a}C\right)
_{\alpha\beta}P_{a}\right]  ,\\
\left\{  Q_{\alpha},\Sigma_{\beta}\right\}   &  =-\frac{1}{2}\left[  \left(
\gamma^{ab}C\right)  _{\alpha\beta}Z_{ab}-2\left(  \gamma^{a}C\right)
_{\alpha\beta}\tilde{Z}_{a}\right]  ,\\
\left\{  \Sigma_{\alpha},\Sigma_{\beta}\right\}   &  =-\frac{1}{2}\left[
\left(  \gamma^{ab}C\right)  _{\alpha\beta}\tilde{Z}_{ab}-2\left(  \gamma
^{a}C\right)  _{\alpha\beta}P_{a}\right]  ,
\end{align}
where we have used the commutation relations of the original superalgebra and
the multiplication law of the semigroup $\left(  \ref{Slaw2}\right)  $. \ The
new superalgebra obtained after a resonant $S$-expansion of $\mathfrak{osp}%
\left(  4|1\right)  $ superalgebra corresponds to a generalized minimal
$AdS$-Lorentz superalgebra in $D=4$.

One can see that a new Majorana spinor charge $\Sigma$ has been introduced as
a direct consequence of the $S$-expansion procedure. \ The introduction of a
second spinorial generator can be found in refs. \cite{AF, Green} in the
supergravity and superstrings context, respectively.

Let us note that a generalized $AdS-$Lorentz algebra $=\left\{  J_{ab}%
,P_{a},\tilde{Z}_{a},\tilde{Z}_{ab},Z_{ab}\right\}  $ forms a bosonic
subalgebra of the new superalgebra and looks very similar to the
$AdS-\mathcal{L}_{6}$ algebra introduced in ref. \cite{SS}. \ In fact one
could identify $\tilde{Z}_{ab}$, $Z_{ab}$ and $\tilde{Z}_{a}$ with
$Z_{ab}^{\left(  1\right)  }$, $Z_{ab}^{\left(  2\right)  }$ and $Z_{a}$ of
$AdS-\mathcal{L}_{6}$ respectively. \ Nevertheless, the commutation relations
$\left(  \ref{dif2}\right)  $ are subtly different of those of the
$AdS-\mathcal{L}_{6}$ algebra.\ On the other hand, the usual $AdS-\mathcal{L}%
_{4}$ algebra $=\left\{  J_{ab},P_{a},Z_{ab}\right\}  $ is also a subalgebra.

It is interesting to observe that an In\"{o}n\"{u}-Wigner contraction of the
new superalgebra leads to a generalization of the minimal Maxwell superalgebra
introduced in ref. \cite{BGKL}. \ After the rescaling%
\begin{align*}
\tilde{Z}_{ab}  &  \rightarrow\mu^{2}\tilde{Z}_{ab},\text{ \ \ }%
Z_{ab}\rightarrow\mu^{4}Z_{ab},\text{ \ \ }P_{a}\rightarrow\mu^{2}P_{a},\text{
\ \ }\\
\tilde{Z}_{a}  &  \rightarrow\mu^{4}\tilde{Z}_{a},\text{ \ \ }Q_{\alpha
}\rightarrow\mu Q_{\alpha}\text{ \ \ and \ \ }\Sigma\rightarrow\mu^{3}%
\Sigma\text{,}%
\end{align*}
the limit $\mu\rightarrow\infty$ provides with a generalized minimal Maxwell
superalgebra $s\mathcal{M}_{4}$ in $D=4$. \ An extensive study of the minimal
Maxwell superalgebra and its generalization has been done using expansion
methods in refs. \cite{AILW, CR1}.\ \ On the other hand, it was shown in refs.
\cite{AI, CR2} that $D=4$, $N=1$ pure supergravity Lagrangian can be obtained
as a quadratic expression in the curvatures associated with the minimal
Maxwell superalgebra.

Analogously to the previous case, we can show that the generalized minimal
$AdS$-Lorentz superalgebra found here can be used in order to build a more
general supergravity action involving a generalized supersymmetric
cosmological term.

As in the previous section, we start from the one-form gauge connection,%
\begin{equation}
A=\frac{1}{2}\omega^{ab}J_{ab}+\frac{1}{l}e^{a}P_{a}+\frac{1}{2}\tilde{k}%
^{ab}\tilde{Z}_{ab}+\frac{1}{2}k^{ab}Z_{ab}+\frac{1}{l}\tilde{h}^{a}\tilde
{Z}_{a}+\frac{1}{\sqrt{l}}\psi^{\alpha}Q_{\alpha}+\frac{1}{\sqrt{l}}%
\xi^{\alpha}\Sigma_{\alpha},
\end{equation}
where the one-form gauge fields are related to the $\mathfrak{osp}\left(
4|1\right)  $ gauge fields $\left(  \tilde{\omega}^{ab},\tilde{e}^{a}%
,\tilde{\psi}\right)  $ as%
\[%
\begin{tabular}
[c]{ll}%
$\omega^{ab}=\lambda_{0}\tilde{\omega}^{ab},$ & $\ e^{a}=\lambda_{2}\tilde
{e}^{a},$\\
$\tilde{k}^{ab}=\lambda_{2}\tilde{\omega}^{ab},$ & $\ \psi^{\alpha}%
=\lambda_{1}\tilde{\psi}^{\alpha},$\\
$k^{ab}=\lambda_{4}\tilde{\omega}^{ab},$ & $\ \xi^{\alpha}=\lambda_{3}%
\tilde{\psi}^{\alpha},$\\
$\tilde{h}^{a}=\lambda_{4}e^{a}.$ &
\end{tabular}
\ \ \ \ \
\]
Then the corresponding two-form curvature $F=dA+A\wedge A$ is given by%
\begin{equation}
F=\frac{1}{2}R^{ab}J_{ab}+\frac{1}{l}R^{a}P_{a}+\frac{1}{2}\tilde{F}%
^{ab}\tilde{Z}_{ab}+\frac{1}{2}F^{ab}Z_{ab}+\frac{1}{l}\tilde{H}^{a}\tilde
{Z}_{a}+\frac{1}{\sqrt{l}}\Psi^{\alpha}Q_{\alpha}+\frac{1}{\sqrt{l}}%
\Xi^{\alpha}\Sigma_{\alpha}, \label{FS2}%
\end{equation}
where%
\begin{align*}
R^{ab}  &  =d\omega^{ab}+\omega_{\text{ }c}^{a}\omega^{cb},\\
R^{a}  &  =de^{a}+\omega_{\text{ }b}^{a}e^{b}+k_{\text{ }b}^{a}e^{b}+\tilde
{k}_{\text{ }b}^{a}\tilde{h}^{b}-\frac{1}{2}\bar{\psi}\gamma^{a}\psi-\frac
{1}{2}\bar{\xi}\gamma^{a}\xi,\\
\tilde{H}^{a}  &  =d\tilde{h}^{a}+\omega_{\text{ }b}^{a}\tilde{h}^{b}%
+\tilde{k}_{\text{ }b}^{a}e^{b}+k_{\text{ }b}^{a}\tilde{h}^{b}-\bar{\psi
}\gamma^{a}\xi,\\
\tilde{F}^{ab}  &  =d\tilde{k}^{ab}+\omega_{\text{ }c}^{a}\tilde{k}%
^{cb}-\omega_{\text{ }c}^{b}\tilde{k}^{ca}+k_{\text{ }c}^{a}\tilde{k}%
^{cb}-k_{\text{ }c}^{b}\tilde{k}^{ca}+\frac{2}{l^{2}}e^{a}\tilde{h}^{b}%
+\frac{1}{2l}\bar{\psi}\gamma^{ab}\psi+\frac{1}{2l}\bar{\xi}\gamma^{ab}\xi,\\
F^{ab}  &  =dk^{ab}+\omega_{\text{ }c}^{a}k^{cb}-\omega_{\text{ }c}^{b}%
k^{ca}+\tilde{k}_{\text{ }c}^{a}\tilde{k}^{cb}+k_{\text{ }c}^{a}k^{cb}%
+\frac{1}{l^{2}}e^{a}e^{b}+\frac{1}{l^{2}}\tilde{h}^{a}\tilde{h}^{b}+\frac
{1}{l}\bar{\xi}\gamma^{ab}\psi,\\
\Psi &  =d\psi+\frac{1}{4}\omega_{ab}\gamma^{ab}\psi+\frac{1}{4}k_{ab}%
\gamma^{ab}\psi+\frac{1}{4}\tilde{k}_{ab}\gamma^{ab}\xi+\frac{1}{2l}%
e_{a}\gamma^{a}\xi+\frac{1}{2}\tilde{h}_{a}\gamma^{a}\psi,\\
\Xi &  =d\xi+\frac{1}{4}\omega_{ab}\gamma^{ab}\xi+\frac{1}{4}k_{ab}\gamma
^{ab}\xi+\frac{1}{4}\tilde{k}_{ab}\gamma^{ab}\psi+\frac{1}{2l}e_{a}\gamma
^{a}\psi+\frac{1}{2l}\tilde{h}_{a}\gamma^{a}\xi.
\end{align*}
Here the new Majorana field $\xi$ is associated to the fermionic generator
$\Sigma$, while the one-forms $\tilde{h}^{a}$, $\tilde{k}^{ab}$ and $k^{ab}$
are the matter fields associated with the bosonic generators $\tilde{Z}_{a}$,
$\tilde{Z}_{ab}$ and $Z_{ab}$ respectively. \

Using the two-form curvature $F$ it is possible to write the action for the
generalized minimal $AdS-$Lorentz superalgebra as%
\begin{equation}
S=2%
{\displaystyle\int}
\left\langle F\wedge F\right\rangle =2%
{\displaystyle\int}
F^{A}\wedge F^{B}\left\langle T_{A}T_{B}\right\rangle _{\mathcal{S}},
\end{equation}
where $\left\langle T_{A}T_{B}\right\rangle _{\mathcal{S}}$ corresponds to an
$S$-expanded invariant tensor which is obtained from the original components
of the invariant tensor $\left(  \ref{invt1}\right)  $. \ \ Using Theorem
VII.1 of ref. \cite{Sexp}, it is possible to show that the non-vanishing
components of $\left\langle T_{A}T_{B}\right\rangle _{\mathcal{S}}$ are given
by%
\begin{align}
\left\langle J_{ab}J_{cd}\right\rangle _{\mathcal{S}}  &  =\alpha
_{0}\left\langle J_{ab}J_{cd}\right\rangle ,\text{ \ \ \ \ \ \ }\left\langle
\tilde{Z}_{ab}\tilde{Z}_{cd}\right\rangle _{\mathcal{S}}=\alpha_{4}%
\left\langle J_{ab}J_{cd}\right\rangle ,\label{invten1}\\
\left\langle J_{ab}\tilde{Z}_{cd}\right\rangle _{\mathcal{S}}  &  =\alpha
_{2}\left\langle J_{ab}J_{cd}\right\rangle ,\text{ \ \ \ \ \ \ }\left\langle
Z_{ab}Z_{cd}\right\rangle _{\mathcal{S}}=\alpha_{4}\left\langle J_{ab}%
J_{cd}\right\rangle ,\\
\text{\ \ \ \ \ \ }\left\langle \tilde{Z}_{ab}Z_{cd}\right\rangle
_{\mathcal{S}}  &  =\alpha_{2}\left\langle J_{ab}J_{cd}\right\rangle ,\text{
\ \ \ \ \ \ }\left\langle J_{ab}Z_{cd}\right\rangle _{\mathcal{S}}=\alpha
_{4}\left\langle J_{ab}J_{cd}\right\rangle ,\\
\left\langle Q_{\alpha}Q_{\beta}\right\rangle _{\mathcal{S}}  &  =\alpha
_{2}\left\langle Q_{\alpha}Q_{\beta}\right\rangle ,\text{ \ \ \ \ \ \ }%
\left\langle \Sigma_{\alpha}\Sigma_{\beta}\right\rangle _{\mathcal{S}}%
=\alpha_{2}\left\langle Q_{\alpha}Q_{\beta}\right\rangle ,\\
\left\langle Q_{\alpha}\Sigma_{\beta}\right\rangle _{\mathcal{S}}  &
=\alpha_{4}\left\langle Q_{\alpha}Q_{\beta}\right\rangle , \label{invten2}%
\end{align}
where $\alpha_{0}$, $\alpha_{2}$ and $\alpha_{4}$ are dimensionless arbitrary
independent constants and
\begin{align*}
\left\langle J_{ab}J_{cd}\right\rangle  &  =\epsilon_{abcd},\\
\left\langle Q_{\alpha}Q_{\beta}\right\rangle  &  =2\left(  \gamma_{5}\right)
_{\alpha\beta}.
\end{align*}

Then considering the two-form curvature $\left(  \ref{FS2}\right)  $ and the
non-vanishing components of the invariant tensor $\left(  \ref{invten1}%
\right)  -\left(  \ref{invten2}\right)  $ we found that the action can be
written as a MacDowell-Mansouri like action,%
\begin{align}
S &  =2\int\left(  \frac{\alpha_{0}}{4}\epsilon_{abcd}R^{ab}R^{cd}%
+\frac{\alpha_{2}}{2}\epsilon_{abcd}R^{ab}\tilde{F}^{cd}+\frac{\alpha_{2}}%
{2}\epsilon_{abcd}\tilde{F}^{ab}F^{cd}+\frac{\alpha_{4}}{2}\epsilon
_{abcd}R^{ab}F^{cd}\right.  \nonumber\\
&  \left.  \frac{\alpha_{4}}{4}\epsilon_{abcd}\tilde{F}^{ab}\tilde{F}%
^{cd}+\frac{\alpha_{4}}{2}\epsilon_{abcd}F^{ab}F^{cd}+\frac{2}{l}\alpha
_{2}\bar{\Psi}\gamma_{5}\Psi+\frac{2}{l}\alpha_{2}\bar{\Xi}\gamma_{5}\Xi
+\frac{4}{l}\alpha_{4}\bar{\Psi}\gamma_{5}\Xi\right)  .\nonumber\\
&
\end{align}
Since we are interested in obtaining the Einstein-Hilbert and the
Rarita-Schwinger like Lagrangian with a generalized supersymmetric
cosmological term, we shall consider only the piece proportional to
$\alpha_{4}$. Using the useful gamma matrix identities and the Bianchi
identities $\left(  dF+\left[  A,F\right]  =0\right)  $ it is possible to
write explicitly the $\alpha_{4}$-term as
\begin{align}
\mathcal{S} &  =\mathcal{\alpha}_{4}\int\epsilon_{abcd}\left(  R^{ab}%
\mathcal{K}^{cd}+\frac{1}{2}\mathcal{\tilde{K}}^{ab}\mathcal{\tilde{K}}%
^{cd}+\frac{1}{2}\mathcal{K}^{ab}\mathcal{K}^{cd}\right)  \nonumber\\
&  +\frac{1}{l^{2}}\left(  \epsilon_{abcd}R^{ab}e^{c}e^{d}+4\bar{\psi}%
e^{a}\gamma_{a}\gamma_{5}D\psi+4\bar{\xi}e^{a}\gamma_{a}\gamma_{5}D\xi\right)
\nonumber\\
&  +\frac{1}{l^{2}}\left(  \epsilon_{abcd}R^{ab}\tilde{h}^{c}\tilde{h}%
^{d}+4\bar{\psi}\tilde{h}^{a}\gamma_{a}\gamma_{5}D\xi+4\bar{\xi}\tilde{h}%
^{a}\gamma_{a}\gamma_{5}D\psi\right)  \nonumber\\
&  +\frac{1}{l^{2}}\epsilon_{abcd}\left(  2\mathcal{\tilde{K}}^{ab}e^{c}%
\tilde{h}^{d}+\mathcal{K}^{ab}e^{c}e^{d}+\mathcal{K}^{ab}\tilde{h}^{c}%
\tilde{h}^{d}+\frac{1}{l^{2}}e^{a}e^{b}e^{c}e^{d}\right.  \nonumber\\
&  \left.  +\frac{6}{l^{2}}e^{a}e^{b}\tilde{h}^{c}\tilde{h}^{d}+\frac{1}%
{l^{2}}\tilde{h}^{a}\tilde{h}^{b}\tilde{h}^{c}\tilde{h}^{d}+\frac{2}{l}%
\bar{\psi}\gamma^{ab}\psi e^{c}\tilde{h}^{d}+\frac{2}{l}\bar{\psi}\gamma
^{ab}\xi e^{c}e^{d}+\frac{2}{l}\bar{\psi}\gamma^{ab}\xi\tilde{h}^{c}\tilde
{h}^{d}\right.  \nonumber\\
&  \left.  +\frac{2}{l}\bar{\xi}\gamma^{ab}\xi e^{c}\tilde{h}^{d}+k^{ab}%
e^{c}\left\{  \bar{\psi}\gamma^{d}\psi+\bar{\xi}\gamma^{d}\xi\right\}
+2\tilde{k}^{ab}e^{c}\bar{\psi}\gamma^{d}\xi+2k^{ab}\tilde{h}^{a}\bar{\psi
}\gamma^{d}\xi\right.  \nonumber\\
&  \left.  +\tilde{k}^{ab}\tilde{h}^{c}\left\{  \bar{\psi}\gamma^{d}\psi
+\bar{\xi}\gamma^{d}\xi\right\}  \right)  +d\left(  \frac{8}{l}\bar{\xi}%
\gamma_{5}\nabla\psi\right)  ,\label{actionsgAdSL4}%
\end{align}
where we have defined%
\begin{align*}
\mathcal{\tilde{K}}^{ab} &  =D\tilde{k}^{cb}+k_{\text{ }c}^{a}\tilde{k}%
^{cb}+k_{\text{ }c}^{b}\tilde{k}^{ac},\\
\mathcal{K}^{ab} &  =Dk^{ca}+\tilde{k}_{\text{ }c}^{a}\tilde{k}^{cb}+k_{\text{
}c}^{a}k^{cb}.
\end{align*}

Here we can see that the first piece corresponds to an Euler invariant term
which can be seen as a Gauss-Bonnet like term and can be written as a boundary
contribution. \ The second piece contains the Einstein-Hilbert term
$\epsilon_{abcd}R^{ab}e^{c}e^{d}$ and the Rarita-Schwinger like Lagrangian.
\ The novelty consists in the contribution of the new spinor field $\xi$ which
is related to the Majorana spinor charge $\Sigma$. \ The fourth
term\ corresponds to a generalized supersymmetric cosmological term built from
the new $AdS$-Lorentz fields. \ The last piece is a boundary term and does not
contribute to the dynamics.

A significant difference with the previous case $\left(  \text{see eq.
}\left(  \ref{actionsAdSL4}\right)  \right)  $ is the presence of the matter
field $\tilde{h}^{a}$ which is related to the new generator $\tilde{Z}_{a}$.
\ In particular, if we consider $\tilde{h}^{a}=0$ and omit boundary
contributions, the term proportional to $\alpha_{4}$ can be written as%
\begin{align}
\mathcal{S} &  =\mathcal{\alpha}_{4}\int\frac{1}{l^{2}}\left(  \epsilon
_{abcd}R^{ab}e^{c}e^{d}+4\bar{\psi}e^{a}\gamma_{a}\gamma_{5}\mathfrak{\nabla
}\psi+4\bar{\xi}e^{a}\gamma_{a}\gamma_{5}\mathfrak{\nabla}\xi\right)
\nonumber\\
&  +\frac{1}{l^{2}}\epsilon_{abcd}\left(  \mathcal{K}^{ab}e^{c}e^{d}+\frac
{1}{l^{2}}e^{a}e^{b}e^{c}e^{d}+\frac{2}{l}\bar{\psi}\gamma^{ab}\xi e^{c}%
e^{d}\right)  ,\label{AdSLfin}%
\end{align}
with%
\begin{align*}
\mathfrak{\nabla}\psi &  =D\psi+\frac{1}{4}k_{ab}\gamma^{ab}\psi+\frac{1}%
{4}\tilde{k}_{ab}\gamma^{ab}\xi,\\
\nabla\xi &  =D\xi+\frac{1}{4}k_{ab}\gamma^{ab}\xi+\frac{1}{4}\tilde{k}%
_{ab}\gamma^{ab}\psi.
\end{align*}
The action $\left(  \ref{AdSLfin}\right)  $ corresponds to a four-dimensional
supergravity action with a generalized supersymmetric cosmological term.
\ Certainly, the choice of bigger semigroups would allow to construct larger
$AdS$-Lorentz superalgebras introducing a more general cosmological term to
supergravity. \ Nevertheless, this procedure would lead to more complicated
actions which we do not consider here.

It is interesting to observe that an In\"{o}n\"{u}-Wigner contraction of the
action $\left(  \ref{AdSLfin}\right)  $ leads us to the $D=4$ pure
supergravity action. \ In fact after the rescaling%
\begin{align*}
\omega_{ab}  &  \rightarrow\omega_{ab},\text{ \ \ }\tilde{k}_{ab}%
\rightarrow\mu^{2}\tilde{k}_{ab},\text{ \ \ }k_{ab}\rightarrow\mu^{4}%
k_{ab},\text{ \ \ }\\
e_{a}  &  \rightarrow\mu^{2}e_{a},\text{ \ \ }\psi\rightarrow\mu\psi\text{
\ \ and \ \ }\xi\rightarrow\mu^{3}\xi\text{, \ \ \ \ }%
\end{align*}
and dividing the action by $\mu^{4}$, the $D=4$ pure supergravity action is
retrieved by taking the limit $\mu\rightarrow\infty$,%
\begin{equation}
\mathcal{S}\mathcal{=\alpha}_{4}\int\frac{1}{l^{2}}\left(  \epsilon
_{abcd}R^{ab}e^{c}e^{d}+4\bar{\psi}e^{a}\gamma_{a}\gamma_{5}D\psi\right)  .
\end{equation}
It was shown in refs. \cite{CR2, AI} that the $N=1$, $D=4$ pure supergravity
can be derived from the minimal Maxwell superalgebra $s\mathcal{M}_{4}$. This
result is not a surprise since the In\"{o}n\"{u}-Wigner contraction of the
generalized minimal $AdS-$Lorentz superalgebra corresponds to the minimal
Maxwell superalgebra $s\mathcal{M}_{4}$.

Let us note that the procedure considered here could be extended to bigger
$AdS$-Lorentz superalgebras whose In\"{o}n\"{u}-Wigner contractions lead to
the Maxwell superalgebras type defined in \cite{CR1}. \ These Maxwell
superalgebras correspond to the supersymmetric extension of the Maxwell
algebras type introduced in refs. \cite{CPRS2, CPRS3}.

\section{Comments and possible developments}

\qquad In the present work we have presented an alternative way of introducing
the supersymmetric cosmological constant to supergravity. Based on the
$AdS$-Lorentz superalgebra we have built the minimal $D=4$ supergravity action
which includes a generalized supersymmetric cosmological constant term. \ For
this purpose we have applied the semigroup expansion method to the
$\mathfrak{osp}\left(  4|1\right)  $ superalgebra allowing us to construct a
MacDowell-Mansouri like action. \ The geometric formulation of the
supergravity theory found here corresponds to a supersymmetric generalization
of the results of refs. \cite{AKL, SS}.

Interestingly, we have shown that the $AdS$-Lorentz superalgebra allows to add
new terms in the supergravity action \`{a} la MacDowell-Mansouri presented in
\cite{CR2}, describing a generalized supersymmetric cosmological constant.
\ In particular, unlike the Maxwell superalgebra, the bosonic fields $k^{ab}$
associated to the generators $Z_{ab}$ appear not only in the boundary terms,
but also in the bulk Lagrangian.

The presence of the bosonic fields $k^{ab}$ in the boundary could be useful in
the context of the duality between string theory realized on an asymptotically
$AdS$ space-time (times a compact manifold M) and the conformal field theory
living on the boundary ( $AdS/CFT$ correspondence) \cite{Maldacena, GKP,
Witten, AGMOO}. \ Interestingly, as shown in ref. \cite{MO}, the introduction
of a topological boundary in a four-dimensional bosonic action is equivalent
to the holographic renormalization procedure in the $AdS/CFT$ context. \ At
the supergravity level, it was shown in ref. \cite{DAA} that the supersymmetry
invariance of the supergravity action is recovered adding appropriate boundary
terms and thus, reproducing the MacDowell-Mansouri action. \ Then, it is
tempting to argue that the presence of the fields $k^{ab}$ in the boundary
would allow not only to recover the supersymmetry invariance in the geometric
approach (rheonomic approach), but also to regularize the supergravity action
in the holographic renormalization context.

We have also presented a more general supergravity action containing a
cosmological constant (see eq. $\left(  \ref{actionsgAdSL4}\right)  $). For
this aim we have introduced the generalized minimal $AdS$-Lorentz superalgebra
using a bigger semigroup. \ This new superalgebra requires the introduction of
a second Majorana spinorial generator $\Sigma$ leading to additional degress
of freedom. \ In particular we have shown that the In\"{o}n\"{u}-Wigner
contraction of the generalized $AdS$-Lorentz superalgebra leads to the minimal
Maxwell superalgebra.

Our results provide one more example of the advantage of the semigroup
expansion method in the geometrical formulation of a supergravity theory.
\ The approach presented here could be useful in order to analyze a possible
extension to higher dimensions. \ Nevertheless, it seems that in odd
dimensions the Chern-Simons (CS) theory is the appropriate formalism in order
to construct a supergravity action. \ For instance, for $D=3$ interesting CS
(super)gravity theories are obtained using the $AdS$-Lorentz (super)symmetries
\cite{DFIMRSV, FISV, HRA}.

On the other hand, it would be interesting to study the extended supergravity
in the geometrical formulation. \ A future work could be analyze the
$N$-extended $AdS$-Lorentz superalgebra introduced in ref. \cite{Sorokas2} and
the construction of $N$-extended supergravities. \ It seems that the semigroup
expansion procedure used here could have an important role in the construction
of matter-supergravity theories.

\bigskip

\section*{Acknowledgements \textbf{\ }}

This work was supported in part by FONDECYT Grants N$%
{{}^\circ}%
$ 1130653. Two of the authors (P.K. C., E.K. R.) authors were supported by
grants from the Comisi\'{o}n Nacional de Investigaci\'{o}n Cient\'{\i}fica y
Tecnol\'{o}gica (CONICYT) and from the Universidad de Concepci\'{o}n, Chile.
\ P.K. C. and E.K. R. wish to thank L. Andrianopoli, R. D'Auria and M.
Trigiante for their kind hospitality at Dipartimieto Scienza Applicata e
Tecnologia (DISAT) of Politecnico di Torino and for enlightening discussions.

\end{document}